\documentclass[twocolumn,pra,superscriptaddress,longbibliography]{revtex4-2}

\usepackage{amsmath,amsfonts,amssymb}
\usepackage{graphicx}
\graphicspath{{figs/}}
\usepackage{float}
\usepackage{xcolor}
\usepackage{physics}
\usepackage{nicefrac}
\usepackage{bbold}
\usepackage[breaklinks=true]{hyperref}
\usepackage{breakcites}
\usepackage[normalem]{ulem}

\usepackage{circledsteps}
\pgfkeys{/csteps/inner color=white}
\pgfkeys{/csteps/outer color=orange}
\pgfkeys{/csteps/fill color=orange}

\newcommand{\ham}{\mathcal{H}}

\newcommand{\matelx}{h_x}
\newcommand{\mately}{h_y}
\newcommand{\matelphi}{h_\varphi}
\newcommand{\infid}{\mathcal{I}}
\newcommand{\infidavg}{\bar{\infid}}
\newcommand{\omegap}{\omega_{1p}}
\newcommand{\omegapp}{\omega_{2p}}
\newcommand{\epsp}[1]{\varepsilon_{#1}}

\newcommand{\epspzero}[1]{\varepsilon_{#1, 0}}

\newcommand{\epspp}{\varepsilon_{2}}
\newcommand{\epsppzero}{\varepsilon_{2, 0}}
\newcommand{\epspptil}{\tilde{\varepsilon}_{2}}
\newcommand{\epspptilzero}{\tilde{\varepsilon}_{2, 0}}

\newcommand{\catp}{\ket{C_\alpha^+}}
\newcommand{\catm}{\ket{C_\alpha^-}}
\newcommand{\catnorm}{\mathcal{N}}

\newcommand{\deltarob}{\delta_{\mathrm{rob}}}

\newcommand{\note}[1]{}
\newcommand{\attention}[1]{}
\newcommand{\draftfig}[1]{}

\newcommand{\AWS}{AWS Center for Quantum Computing, Pasadena, CA 91125, USA}
\newcommand{\UChicago}{Department of Computer Science, University of Chicago, Chicago, Illinois 60637, USA}

\begin{document}

\title{Frequency-noise-insensitive universal control of Kerr-cat qubits}

\author{Lennart Maximilian Seifert}
\affiliation{\UChicago}

\author{Connor T.~Hann}
\affiliation{\AWS}

\author{Kyungjoo Noh}
\affiliation{\AWS}

\date{\today}

\begin{abstract}
    We theoretically study the influence of frequency uncertainties on the operation of a Kerr-cat qubit. As the mean photon number increases, Kerr-cat qubits provide an increasing level of protection against phase errors induced by unknown frequency shifts during idling and $X$ rotations. However, realizing rotations about the other principal axes (e.g., $Y$ and $Z$ axes) while preserving robustness is nontrivial.
    To address this challenge, we propose a universal set of gate schemes which circumvents the tradeoff between protection and controllability in Kerr-cat qubits and retains robustness to unknown frequency shifts to at least first order. Assuming an effective Kerr oscillator model, we theoretically and numerically analyze the robustness of elementary gates on Kerr-cat qubits, with special focus on gates along nontrivial rotation axes. An appealing application of this qubit design would include tunable superconducting platforms, where the induced protection against frequency noise would allow for a more flexible choice of operating point and thus the potential mitigation of the impact of spurious two-level systems.
\end{abstract}

\maketitle

To build a useful quantum computer which can solve problems of practical importance, it is crucial to increase the size of quantum computing systems while ensuring sufficiently low component error rates. In recent years, the field of quantum computing has witnessed significant progress as larger systems with lower error rates have been realized with various hardware platforms \cite{bluvsteinLogicalQuantumProcessor2024,acharyaQuantumErrorCorrection2024,puttermanHardwareefficientQuantumError2025,paetznickDemonstrationLogicalQubits2024,wangSurfaceParticipationDielectric2015,mullerUnderstandingTwolevelsystemsAmorphous2019}. Despite the rapid progress, however, it remains a challenge to produce a large quantum computing system in a reliable manner. For example, performance of a large-scale superconducting quantum device is often limited by fabrication disorders and spurious two-level systems (TLSs). In particular, TLSs often move close to a subset of qubits in an unpredictable manner and significantly degrade the performance of the affected qubits (e.g., substantially reducing qubit lifetimes) \cite{degraafTwolevelSystemsSuperconducting2020, klimovFluctuationsEnergyRelaxationTimes2018,youStabilizingImprovingQubit2022,choSimulatingNoiseQuantum2023,martinisDecoherenceJosephsonQubits2005}. 

\begin{figure}[t!]
    \centering
    \includegraphics[scale=1.]{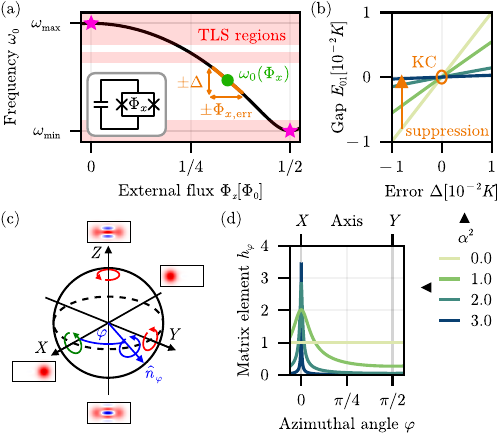}
    \caption{
        (a) In frequency-tunable transmons (inset), uncertainties in the flux $\Phi_{x, \text{err}}$ lead to unknown frequency shifts $\Delta$ (orange arrows around green point) when operated away from sweet spots (pink stars). (b) Shifts $\Delta$ break the degeneracy of the KC qubit (orange circle), causing phase errors. The gap opening is exponentially suppressed with the cat size $\alpha^2$, creating robustness to $\Delta$. (c) Bloch sphere of the KC qubit showing Wigner functions of the $X$ and $Z$ eigenstates as well as rotations about the axes $X$ (easy, green), $Y$ and $Z$ (hard, red), and an arbitrary axis in the $X$-$Y$ plane $\hat{n}_\varphi$ (blue). (d) Dependence of the magnitude of the drive matrix element $\matelphi$ on the angle $\varphi$. As the cat size and protection increase, rotations about all axes in the $X$-$Y$ plane except for the $X$ axis get suppressed.}
    \label{fig:kc}
\end{figure}

The negative effects of TLSs are often mitigated by using frequency-tunable transmons \cite{kochChargeinsensitiveQubitDesign2007, schreierSuppressingChargeNoise2008,barendsSuperconductingQuantumCircuits2014} and operating them at a frequency that is off-resonant with any nearby TLSs. However, this might require tuning the qubit away from its ``sweet spot'', where the frequency is insensitive to its control knob to first order, as illustrated in Fig.~\ref{fig:kc}(a). Thus, reduction in the TLS-induced errors on a frequency-tunable qubit comes at the cost of increased frequency noise and dephasing caused by undesired fluctuations in the control knob, without any protection from a first-order insensitivity.

Taking inspiration from this example, in this letter, we propose a general approach for robustly operating qubits encoded in Kerr-nonlinear oscillators that are subject to undesired fluctuations in their frequency (e.g., frequency-tunable transmons). Our proposal is based on Kerr-cat qubits (KCs) \cite{cochraneMacroscopicallyDistinctQuantum1999, grimmStabilizationOperationKerrcat2020,puriEngineeringQuantumStates2017,puriStabilizedCatDriven2019,xuEngineeringFastBiaspreserving2022,ruizTwophotonDrivenKerr2023,hajrHighCoherenceKerrCatQubit2024,aokiControlCouplingKerr2024,venkatramanNonlinearDissipationDriven2024,venkatramanDrivenKerrOscillator2024, garcia-mataEffectiveFloquetTheory2024,darmawanPracticalQuantumError2021,iyamaObservationManipulationQuantum2024,frattiniObservationPairwiseLevel2024,puriBiaspreservingGatesStabilized2020,gravinaCriticalSchrodingerCat2023,gotoUniversalQuantumComputation2016}, which can suppress dephasing errors due to unknown frequency shifts by increasing the mean photon number of the qubit states. Importantly, we address the challenges associated with reduced controllability of KCs in the protected regime. In particular, we propose gate schemes that overcome the tradeoff between protection and controllability and enable universal computation with KCs, while being highly robust to dephasing errors that emerge from undesired frequency noise. Thus our work is driven by a fundamentally different motivation compared to past studies on KCs, which typically focused on engineering a strong noise bias through large mean photon numbers and realizing bias-preserving gates \cite{xuEngineeringFastBiaspreserving2022, puriBiaspreservingGatesStabilized2020,puriStabilizedCatDriven2019}. Here we specifically consider the regime of smaller mean photon numbers ($\bar{n} \lesssim 3$) in order to limit the negative impact of increasing bit-flip rates ($\propto \bar{n}$) in KCs. We further note that, compared to previous approaches to realizing frequency-noise-insensitive control of superconducting qubits (e.g., via spin-locking \cite{zukRobustGatesSpinlocked2024}), our approach allows for always-on frequency noise protection across a universal set of single- and two-qubit gates.


\textit{Protection versus controllability of the Kerr-cat}---We consider a Kerr-nonlinear oscillator subject to a two-photon squeezing drive with frequency $\omegapp$. In the frame rotating with $\omegapp/2$, the system Hamiltonian reads
\begin{equation}
    \label{eq:ham_drift}
    \ham = \delta a^\dagger a - \frac{K}{2} a^{\dagger 2}a^2 + \frac{\epspp}{2} \qty(a^2 + a^{\dagger 2}),
\end{equation}
where $\delta = \omega_0 - \omegapp/2$ is the detuning between frequency of the bare qubit $\omega_0$ and the rotating frame,  $\epspp$ is the two-photon drive strength, and $K$ is the Kerr nonlinearity. Without loss of generality we assume $K > 0$.

Eq.~\eqref{eq:ham_drift} shows that the system has a symmetry in the photon number parity, therefore the eigenstates are superpositions of either only even or only odd photon number states. We denote by $\ket{2k}$ the $(k+1)$th even parity state and by $\ket{2k+1}$ the $(k+1)$th odd parity state, $k \geq 0$, sorted by energy. The computational subspace is spanned by the states $\ket{0}$ and $\ket{1}$ with energy gap $E_{01}(\delta, \alpha^2) = E_1 - E_0$.
On resonance at $\delta = 0$, these states are degenerate ($E_{01} = 0$) and furthermore correspond to Schrödinger cat states $\ket{0} = \catp = (\ket{\alpha} + \ket{-\alpha})/\catnorm_+$ and $\ket{1} = \catm = (\ket{\alpha} - \ket{-\alpha})/\catnorm_-$ with normalization constants $\catnorm_\pm = \sqrt{2(1 \pm \exp(-2\alpha^2))}$. This defines the standard KC with cat size $\alpha^2 = \epspp / K \ge 0$ and mean photon number $\bar{n} \simeq \alpha^2$. 

Unknown shifts $\Delta$ in the oscillator's frequency break the KC's degenerate qubit manifold in the basis $\qty{\ket{0}, \ket{1}}$ and open the energy gap to first order like $E_{01}(\Delta, \alpha^2) \simeq \partial_\delta E_{01}(0, \alpha^2) \, \Delta$, leading to the accumulation of undesired phase errors. However, the appeal of the KC is that such phase errors are exponentially suppressed with the cat size via $\partial_\delta E_{01}(0, \alpha^2) \simeq 4 \alpha^2 \exp(-2 \alpha^2)$. 
Fig.~\ref{fig:kc}(b) visualizes the energy gap $E_{01}(\Delta, \alpha^2)$ as a function of the frequency uncertainty $\Delta$, which shows the exponential suppression as the cat size $\alpha^2$ increases. This illustrates the inherent protection of KCs to phase errors induced by unknown frequency shifts when idling. In the remainder of this article we address the question of whether this robustness can be retained while simultaneously realizing universal control of the KC. 

The key challenge arises from the difficulty to implement arbitrary single-qubit gates. While the inherent protection mitigates unwanted perturbations, it also limits the controllability. For example, rotations about axes in the $X$-$Y$ plane are realized by applying a single-photon drive with frequency $\omegap = \omegapp/2$, which adds the Hamiltonian term $\varepsilon(t) \qty(e^{i \tilde{\varphi}} a + e^{-i \tilde{\varphi}} a^\dagger)/2$. The relevant matrix element for such KC rotations is given by
\begin{equation}
    \matelphi e^{i \varphi} \equiv \mel{1}{\qty(e^{i \tilde{\varphi}} a + e^{-i \tilde{\varphi}} a^\dagger)}{0},
\end{equation}
where $\hat{n}_\varphi = (\cos(\varphi), \sin(\varphi), 0)^\intercal$ describes the rotation axis in the Bloch sphere. Due to the KC protection mechanism the relationship between the drive phase $\tilde{\varphi}$ and the azimuthal angle $\varphi$ is nontrivial; we include the derivation in the Supplemental Information (SI) \cite{SupplementalInformation}. The magnitude $\matelphi$ determines the achievable Rabi rate $\Omega \propto \matelphi$ for a specific axis and we show its dependence on $\varphi$ in Fig.~\ref{fig:kc}(c). As the protection of the KC increases, i.e. $\alpha^2$ gets larger, rotations about all axes in the $X$-$Y$ plane except for the $X$ axis are suppressed as a consequence. In the following we analyze the impact of this imbalance on the performance of rotations about the principal axes $X$ ($\varphi = 0$, $\matelphi \rightarrow \matelx$) and $Y$ ($\varphi = \pi/2$, $\matelphi \rightarrow \mately$).


\textit{$X$ rotations}---Fig.~\ref{fig:kc}(c) underlines the well-known fact that for large cat sizes only $X$ rotations are not suppressed \cite{grimmStabilizationOperationKerrcat2020,hajrHighCoherenceKerrCatQubit2024,frattiniObservationPairwiseLevel2024}. In fact, the relevant matrix element grows like $\matelx \simeq 2 \alpha$. This enables the straightforward realization of arbitrary $X(\theta)$ gates under the KC protection, so we refer to $X$ as the ``easy'' rotation axis. 

We validate the robustness of a $X\qty(\pi/2)$ rotation (target gate $V$) to frequency noise by simulating the dynamics (actual gate $U(T, \Delta)$) and computing the infidelity
\begin{equation}
    \label{eq:infid}
    \infid(T, \Delta) = 1 - \frac{1}{4} \abs{\Tr[V^\dagger P U(T,\Delta) P]\!}^2
\end{equation}
for gate times $T \in \qty[10 \, K^{-1}, 50 \, K^{-1}]$ and frequency shifts $\Delta \in \qty[-\Delta_\textrm{max}, \Delta_\textrm{max}]$ with $\Delta_\mathrm{max} = 5 \cdot 10^{-3} \, K$. $P$ is the projection to the computational subspace, thus $\infid(T, \Delta)$ also captures leakage errors. Were there no protection, this choice for $T$ and $\Delta_\mathrm{max}$ would correspond to coherent errors in the range $(\Delta_\mathrm{max} T)^2 \sim 10^{-3} - 10^{-1}$.  In the case of superconducting platforms with typical Kerr values $K/2\pi \sim 10 - 100~\mathrm{MHz}$, we would have characteristic time units of $K^{-1} \sim 16 - 1.6~\mathrm{ns}$ and frequency shifts up to $\Delta_\mathrm{max}/2\pi \sim 50 - 500~\mathrm{kHz}$.
In the main text we focus on static shifts $\Delta$ which allow for conclusions about errors with low-frequency signatures (``quasistatic'') and we discuss time-dependent errors $\Delta(t)$ in the SI \cite{SupplementalInformation}. 

\begin{figure*}[htbp]
    \centering
    \includegraphics[scale=1.]{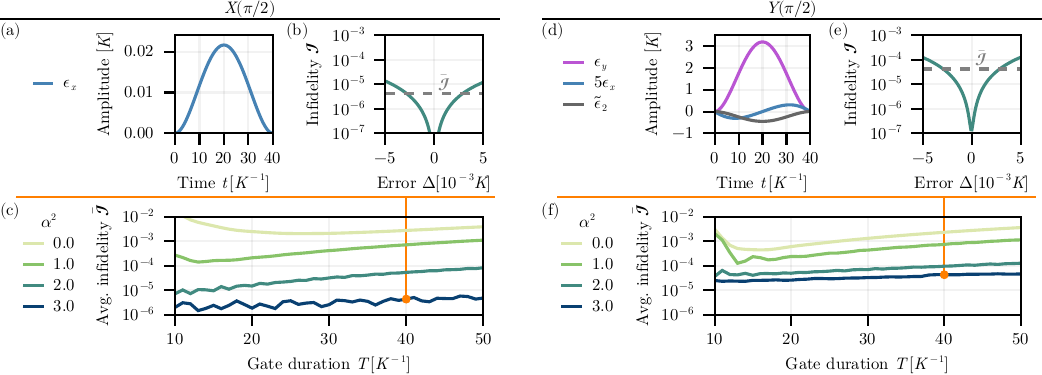}
    \caption{Pulse implementation and infidelity robustness for the $X(\pi/2)$ and $Y(\pi/2)$ gates. (a) The $X$ rotation is native to the KC and is realized through a simple single-photon drive pulse. (b) The infidelity of the pulse (a) when simulated over a range of detuning errors $\Delta$. (c) The average infidelity $\infidavg$ of pulses of type (a) for different durations and cat sizes. The infidelity decreases when the cat size is increased. (d) Pulse scheme for the $Y$ rotation, which includes tuning the two-photon the pump strength $\epspptil$ to lower the cat size and a DRAG term $\epsilon_x$. (e) The infidelity of the pulse (d) when simulated over a range of detuning errors $\Delta$. (f) The average infidelity $\infidavg$ of pulses of type (d) for different durations and cat sizes. The increase in robustness diminishes when increasing the cat size beyond $\alpha^2 \gtrsim 2$.}
    \label{fig:xy_gate}
\end{figure*}

For the pulse shape $\epsp{x}(t)$ we consider the truncated Gaussian that was used in Ref.~\cite{xuEngineeringFastBiaspreserving2022}. Given a gate time $T$, we find the optimal drive amplitude that minimizes the average infidelity 
\begin{equation}
    \label{eq:avg_infid}
    \infidavg(T) = \frac{1}{2\Delta_\mathrm{max}} \int_{-\Delta_\mathrm{max}}^{\Delta_\mathrm{max}} \dd{\Delta} \infid(T, \Delta).
\end{equation}
In Fig.~\ref{fig:xy_gate}(a) we graph the pulse shape for $T=40~K^{-1}$, which achieves infidelities $\infid \lesssim 10^{-4}$ over the $\Delta$ range as shown in Fig.~\ref{fig:xy_gate}(b). This constitutes one data point in Fig.~\ref{fig:xy_gate}(c), where we show the average infidelity $\infidavg$ over the range of gate durations. We observe that $\infidavg$ decreases (the robustness increases) for greater $\alpha^2$, as expected, reaching values orders of magnitude below the naive coherent error limit $(\Delta_\text{max} T)^2$ thanks to the KC's robustness. We note that infidelities increase with larger gate time as the phase errors are not perfectly eliminated and accumulate more over longer durations. The comparatively large infidelity in the short gate time regime for small cat sizes is because the required drive strength becomes comparable to the energy gap to excited states, leading to leakage.


\textit{Two-qubit gate}---Before we turn to the more challenging single-qubit gates (the main focus of this work), we briefly discuss the realization of a two-qubit gate that is analogous to the $X(\theta)$ gate. We consider two KC qubits $i = A, B$ coupled via a beamsplitter interaction \cite{aokiResidualZZcouplingSuppressionFast2024,puriBiaspreservingGatesStabilized2020}. On a superconducting architecture this could be realized by driving a tunable transmon coupler \cite{aokiControlCouplingKerr2024} or SNAIL coupler \cite{chapmanHighOnOffRatioBeamSplitterInteraction2023} at the difference frequency $\abs{\omega_0^A - \omega_0^B}$. In the KC basis the effective Hamiltonian reads \cite{SupplementalInformation}
\begin{equation}
    \label{eq:ham_int}
    \ham_\mathrm{int}(t) = \frac{g(t)}{2} \qty(\matelx^A \matelx^B X_A X_B + \mately^A \mately^B Y_A Y_B).
\end{equation}
For the bare qubit with $\alpha^2 = 0$ this reduces to an $\mathrm{iSWAP}$ interaction. In the KC case we instead target an $XX(\theta)$ gate as the vanishing $\mately^i \propto e^{-2 \alpha_i^2}$ elements \cite{SupplementalInformation} suppress the $YY$ term. However, since the $YY$ term may be non-negligible in the regime of smaller $\alpha^2$, an echo sequence
\begin{equation}
    XX(\theta) = X_i R_\mathrm{int}(\theta/2) X_i R_\mathrm{int}(\theta/2)
\end{equation}
can be employed to fully cancel the $YY$ term. Here either $i=A$ or $i=B$ is possible and $R_\mathrm{int}(\eta)$ is the rotation generated by $\ham_\mathrm{int}$ with angle $\eta = \int_0^T \dd{t} g(t) \matelx^A \matelx^B$. The special choice $XX(\pi/2)$ is equivalent to the Mølmer-Sørensen gate on trapped-ion platforms \cite{sorensenQuantumComputationIons1999}, which is locally equivalent to  $\mathrm{CNOT}$ and $\mathrm{CZ}$. Thus, similarly to the $X(\theta)$ case, an entangling gate can be directly realized in a robust way as the KCs have an exponentially suppressed sensitivity to detuning-induced dephasing while offering a large matrix element product $\matelx^A \matelx^B \simeq 4 \alpha_A \alpha_B$ to drive the interaction.

\textit{$Y$ rotations}---In order to realize a universal gate set, single-qubit rotations about at least one ``nontrivial'' axis are required. A natural first candidate are $Y(\theta)$ gates as they can also be realized with a single-photon drive, similarly to $X(\theta)$ gates. However, $Y$ rotations are exponentially suppressed in the KC (see Fig.~\ref{fig:kc}(c)). 
To overcome this issue, we temporarily reduce the cat size $\alpha^2(t)$ by adiabatically ramping down the two-photon drive amplitude $\epspp(t) = \epsppzero + \epspptil(t)$, where $\epspptil(t) \leq 0$. This trades off an increased matrix element $\mately(t)$ for smaller inherent protection, which we expect to be partially compensated by a dynamical robustness effect emerging from the gate drive itself, similar to a spin-locking effect \cite{zukRobustGatesSpinlocked2024}. Both the $Y$ gate drive $\epsilon_y(t)$ as well as the ramping $\epspptil(t)$ are given by truncated Gaussians (with optimizable amplitudes $\epspzero{y}$ and $\epspptilzero$), and we further include a corrective component $\epsilon_x(t)$ inspired by the Derivate Removal by Adiabatic Gate (DRAG) scheme \cite{motzoiSimplePulsesElimination2009,gambettaAnalyticControlMethods2011}. We refer to the SI \cite{SupplementalInformation} for more details on the pulse design and parameter optimization. 

\begin{figure*}[htbp]
    \centering  
    \includegraphics[scale=1.]{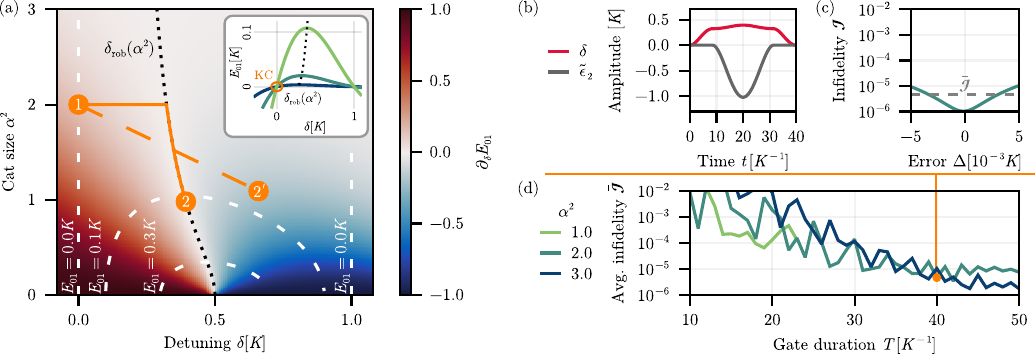}
    \caption{Adiabatic realization of the $Z(-\pi/2)$ gate using the scheme \Circled{1}$\leftrightarrow$\Circled{2}. (a) The energy gap derivative $\partial_\delta{E_{01}}(\delta, \alpha^2)$ as a function of the detuning $\delta$ and cat size $\alpha^2$. The robust line $\deltarob(\alpha^2)$ is highlighted by a black dashed line. An example gate trajectory is shown with an orange solid line. The inset (same legend as (d)) shows the robust line in the energy spectrum. (b) Control pulse for the solid trajectory shown in (a), which simultaneously tunes the detuning $\delta$ and the two-photon pump strength via $\epspptil$. (c) The infidelity of the pulse (b) when simulated over a range of detuning errors $\Delta$. (d) The average infidelity $\infidavg$ of pulses of type (b) for different durations and cat sizes. We note that the $\alpha^2 = 1$ line ends at $T \approx 22~K^{-1}$ because our scheme \Circled{1}$\leftrightarrow$\Circled{2} is not applicable in the case of small cat sizes ($\alpha^2 \lesssim 1$), which is also why $\alpha^2 = 0$ is excluded here. The SI \cite{SupplementalInformation} discusses the alternative scheme \Circled{1}$\leftrightarrow$\Circled{2'}, which works for all cat sizes but is less suited for dynamical errors $\Delta(t)$.}
    \label{fig:z_gate}
\end{figure*}

Figs.~\ref{fig:xy_gate}(d)-(f) present the results of the robustness study for $Y(\pi/2)$.
We note that, while the robustness generally grows with $\alpha^2$ and improves over the spin-locking-only case at $\alpha^2 = 0$, the average infidelity $\infidavg$ does not decrease significantly as $\alpha^2 = 2 \rightarrow 3$. This is in stark contrast to the $X(\pi/2)$ case where the exponentially growing robustness is evident, which is because the KC protection has to be partially sacrificed in order to realize the $Y(\pi/2)$ gate when starting with a greater cat size, and the spin-locking effect of the gate drive provides only limited robustness. While this scheme can produce average infidelities well below $10^{-4}$ on short timescales for these intermediate cat sizes, it requires possibly prohibitively large drive amplitudes $\epspzero{y}$ (see SI \cite{SupplementalInformation}). In what follows, we present an alternative gate scheme for a ``nontrivial'' axis and show that a greater level of robustness can be achieved by harnessing the structure of KC qubit's energy spectrum, going beyond the limitation of a spin-locking mechanism while also avoiding the necessity for strong gate drives.

\textit{$Z$ rotations}---The $Z$ axis is the other ``nontrivial'' principal axis as rotations are directly suppressed by the KC's phase error protection. Prior works \cite{puriEngineeringQuantumStates2017,gotoUniversalQuantumComputation2016,kanaoQuantumGateKerr2022} suggested intentionally detuning the KC ($\delta(t) \neq 0$). However, due to the KC protection mechanism and the resulting suppressed energy gap $E_{01}(\delta, \alpha^2)$ (see Fig. \ref{fig:kc}(b) and inset of Fig. \ref{fig:z_gate}), this leads to gate times that are exponentially slow in $\alpha^{2}$, which is impractical for larger cat sizes. To overcome this controllability issue, we consider an approach where the cat size is adjusted during the $Z(\theta)$ gate in addtion to the detuning. In particular, we propose an adiabatic $Z(\theta)$ scheme that can be implemented on a similar timescale as $X(\theta)$ gates while retaining robustness to first order, thereby circumventing the usual tradeoff between protection and controllability. 


Our scheme relies on the key observation that for a detuned KC there exists a ``robust'' line $\deltarob(\alpha^2) \in [0, K]$ where the energy gap $E_{01}(\deltarob(\alpha^2), \alpha^2) > 0$ is positive but also first-order insensitive to detuning errors, i.e. $\partial_\delta E_{01} (\deltarob, \alpha^2) = 0$. This is visualized in Fig.~\ref{fig:z_gate}(a) with black dotted lines. The solid orange line highlights an example gate trajectory of our proprosed $Z(\theta)$ scheme, which works as follows:
Starting from the initial KC with size $\alpha^2$ (\Circled{1}), we adiabatically change the detuning $\delta = 0 \rightarrow \deltarob(\alpha_1^2)$ over a time $\tau$. Next, over a time $(T-\tau)/2$, we adiabatically trace the robust line by ramping down the cat size $\alpha^2 \rightarrow \alpha'^2 < \alpha^2$ while adjusting the detuning $\delta(t) = \deltarob(\alpha^2(t))$ (midpoint \Circled{2}, cat size $\alpha'^2$). This increases the energy gap $E_{01}$ to accrue a logical phase difference more quickly while the system remains first-order insensitive to unwanted detuning errors $\Delta$. 
Finally, we reverse the process to complete the gate (\Circled{1}). Fig.~\ref{fig:z_gate}(a) shows how the gate trajectory is confined to regions of parameter space where $\partial_\delta E_{01} \ll 1$, suggesting a gate with high robustness and average fidelity overall.

We perform the robustness study for a $Z(-\pi/2)$ gate following this scheme. The negative rotation angle is due to the nonnegative energy gap $E_{01} \geq 0$. The detuning ramps  $\delta = 0 \leftrightarrow \deltarob$ as well as the pump modulation $\epspptil(t)$ are based on truncated Gaussians, and the optimizable parameters are the ramp time $\tau$ and the amplitude $\epspptilzero$ \cite{SupplementalInformation}. An example pulse is included in Fig.~\ref{fig:z_gate}(b). 
We observe in Figs.~\ref{fig:z_gate}(c)-(d) that this gate scheme can achieve average infidelities $\infidavg$ in the $10^{-4} - 10^{-6}$ range, which means similar performance to native $X(\theta)$ rotations at gate times that are within the same order of magnitude (tens of $K^{-1}$). This underlines that utilizing the robust line $\deltarob(\alpha^2)$ is crucial for realizing a protected gate. 

While this scheme \Circled{1}$\leftrightarrow$\Circled{2} is also useful for time-dependent shifts $\Delta(t)$ \cite{SupplementalInformation}, we further identify an alternative scheme \Circled{1}$\leftrightarrow$\Circled{2'} specifically for static errors $\Delta$. We first note that the gate angle $\theta$ is approximately given by the integrated energy gap over the gate trajectory, which to first order in $\Delta$ reads
\begin{equation}
    \label{eq:gate_angle}
    \theta \simeq - \int_0^T \dd{t} E_{01}(t; \Delta = 0) - \Delta \int_0^T \dd{t} \partial_\delta E_{01}(t; \Delta = 0).
\end{equation}
Thus, we observe that first-order robustness to $\Delta$ emerges when the energy gap derivative $\partial_\delta E_{01}$ averages out over the gate evolution. It is characteristic for gate trajectories for this scheme that they cross over the robust line $\deltarob(\alpha^2)$ (see orange dashed line in Fig.~\ref{fig:z_gate}(a)) due to a change in sign of the gap derivative. While the \Circled{1}$\leftrightarrow$\Circled{2'} is only practical for a smaller class of errors, its appeal is in a simpler experimental implementation: The robust line does not have to be calibrated and it is sufficient to optimize a small number of pulse parameters to eliminate the second integral term in Eq. \eqref{eq:gate_angle}. We include the full discussion and analysis in the SI \cite{SupplementalInformation}.

Due to the retained protection, our schemes offer a better error scaling than the fundamentally different Kerr gate scheme, which implements a discrete $Z(\pi/2)$ rotation and has been demonstrated in experiment \cite{grimmStabilizationOperationKerrcat2020,frattiniObservationPairwiseLevel2024}. That approach involves rapidly turning off the two-photon pump ($\epspp = 0$) and letting the system evolve under the Kerr Hamiltonian for a time $T = \pi/K$. During the gate evolution the qubit is fully unprotected and $\infid \simeq \alpha^2 \Delta^2 T^2$ for $\alpha^2 \gtrsim 1$, thus the error grows with cat size. Our schemes enable $\infid = \mathcal{O}(\Delta^4 T^2 K^{-2})$ as they cancel the lowest-order error term.
This comes at the cost of gate times $T = \pi/(2\bar{E}_{01})$ (for $Z(-\pi/2)$) that are about an order of magnitude longer than for the Kerr gate ($\bar{E}_{01} \sim 0.1 \, K$ is the energy gap averaged over the gate trajectory, see Fig.~\ref{fig:z_gate}(a)).


\textit{Discussion}---We introduced a robust approach to universally control Kerr-cat qubits (KCs) with resiliency to frequency fluctuations, which circumvents the KC's usual tradeoff of inherent protection at the cost of reduced controllability. We discuss gate schemes for rotations about the principal axes $X$, $Y$ and $Z$ in the Bloch sphere as well as the two-qubit $XX$ rotation and analyze their robustness against frequency shifts $\Delta$. The ``easy'' gates $X(\theta)$ and $XX(\theta)$ are native to the KC as they can be directly implemented for arbitrary cat size and therefore are trivially robust due to the exponential KC protection. While $Y(\theta)$ rotations can be realized with low infidelity (to a certain limit) but require large drive power, our carefully designed $Z(\theta)$ scheme demonstrates high robustness without facing any physical barriers. It leverages tracing an error-insensitive region $\deltarob(\alpha^2)$ in parameter space $(\delta, \alpha^2)$, which makes the gate robust throughout the entire state evolution.

This ability to perform arbitrary rotations about the axes $\qty{X, Z, XX}$ suffices for universal computation, motivating the concept of a robust KC-based platform. For instance, in the context of tunable superconducting hardware, this could enable the design of a TLS-resistant processor where individual physical devices are operated away from their sweet spots. Alternatively, one could imagine a hybrid platform where devices are used as transmons at their sweet spots ($\alpha^2 = 0$) by default and are only tuned away and converted to KCs ($\alpha^2 > 0$) when an interfering TLS is observed. Note that $XX$ rotations are natively possible for any pair of cat sizes (see Eq.~\eqref{eq:ham_int}), which allows to directly interact a bare transmon with a KC. This hybrid approach would also reduce the impact of increased bit-flip rates in KCs. Lastly, we remark that since our work provides strategies for circumventing the usual tradeoff between protection and controllability in protected qubits, it opens up interesting future research directions for enhancing the controllability of various types of protected qubits such as fluxonium qubits and $0-\pi$ qubits without significantly compromising the protection.     






\textit{Acknowledgments}---We thank Akshay Koottandavida for helpful feedback on the manuscript. The majority of the work was completed as part of L.M.S.'s internship at the AWS Center for Quantum Computing. We thank Simone Severini, James Hamilton, Nafea Bshara, and Peter DeSantis at AWS, for their involvement and support of the research activities at the AWS Center for Quantum Computing. This work is funded in part by the STAQ project under award NSF Phy-232580; in part by the US Department of Energy Office of Advanced Scientific Computing Research, Accelerated Research for Quantum Computing Program.

\bibliographystyle{naturemag}
\bibliography{arxiv}

\pagebreak
\widetext
\begin{center}
\textbf{\large{Frequency-noise-insensitive universal control of Kerr-cat qubits\\Supplemental Information}}
\end{center}
\setcounter{equation}{0}
\setcounter{figure}{0}
\setcounter{table}{0}
\makeatletter
\renewcommand{\theequation}{S\arabic{equation}}
\renewcommand{\thefigure}{S\arabic{figure}}

\section{Analytical expressions in the Kerr-cat computational subspace}
\subsection{Single-photon drive}
\label{sec:1p}
We provide the analytical derivation that highlights the exponential suppression of rotations about all axes except for the $\pm X$ axis in the Bloch sphere. In the computational basis $\qty{\catp \equiv \ket{0}, \catm \equiv \ket{1}}$, the single-photon drive amplitude $\varepsilon(t)$ and phase $\tilde{\varphi}$ becomes
\begin{equation}
    \frac{\varepsilon(t)}{2} \qty(e^{i \tilde{\varphi}} a + e^{-i \tilde{\varphi}} a^\dagger) \rightarrow \frac{\varepsilon(t)}{2} \qty(\cos(\tilde{\varphi}) \matelx X - \sin(\tilde{\varphi}) \mately Y),
\end{equation}
because the ladder operator can be written as $a \rightarrow (\matelx X + i \mately Y) / 2$, where
\begin{equation}
    \matelx = \frac{2 \alpha}{\sqrt{1 - e^{-4 \alpha^2}}}, \qquad \mately = \frac{2 \alpha e^{-2 \alpha^2}}{\sqrt{1 - e^{-4 \alpha^2}}} = \matelx e^{-2 \alpha^2}.
\end{equation}
The relevant transition matrix element is given by
\begin{equation}
    \label{eq:matel}
    \matelphi e^{i \varphi} \equiv \mel{1}{\qty(\cos(\tilde{\varphi}) \matelx X - \sin(\tilde{\varphi}) \mately Y)}{0} = \matelx \qty(\cos(\tilde{\varphi}) - i \sin(\tilde{\varphi}) e^{-2 \alpha^2}).  
\end{equation}
This already hints at the exponential suppression with the cat size $\alpha^2$ when $\tilde{\varphi} \neq m \pi$, $m \in \mathbb{N}$. The rotation axis $\hat{n}_\varphi = (\cos(\varphi), \sin(\theta), 0)$ in the Bloch sphere is determined by the azimuthal angle $\varphi$, which is related to the drive phase due to Eq.~\eqref{eq:matel}:
\begin{gather}
    \varphi = \arg\qty(\cos(\tilde{\varphi}) - i \sin(\tilde{\varphi}) e^{-2 \alpha^2}) \\
    \label{eq:phiphi}
    \iff \sin(\varphi) \cos(\tilde{\varphi}) = - \sin(\tilde{\varphi}) \cos(\varphi) e^{-2 \alpha^2}.
\end{gather}
We obtain the magnitude of the matrix element $\matelphi$ as a function of $\varphi$, which is shown in Fig.~1(c) of the main text:
\begin{align}
    \matelphi &= \matelx \sqrt{\cos^2(\tilde{\varphi}) + \sin^2(\tilde{\varphi})e^{-4\alpha^2}} \\
    &=\matelx \frac{e^{-2\alpha^2}}{\sqrt{1 - \cos^2(\varphi) \qty(1 - e^{-4 \alpha^2})}} \\[5pt]
    &\label{eq:matelmag}
    =\matelx \begin{cases}
        \quad 1 - 2 \alpha^2 \sin^2(\varphi) & \qquad \text{for } \alpha^2 \ll 1,\\[10pt]
        \begin{cases}
            \, 1 & \text{for $\varphi = m\pi$ ($\pm X$ axis)},\\
            \, \sqrt{1 - \cos^2(\varphi)}^{-1} e^{-2\alpha^2} + \mathcal{O}\qty(e^{-4\alpha^2}) & \text{otherwise},
        \end{cases} & \qquad \text{for } \alpha^2 \gg 1.
    \end{cases}
\end{align}
The second line is derived using the square of Eq.~\eqref{eq:phiphi} and the Pythagorean trigometric identity. Eq.~\eqref{eq:matelmag} shows the exponential suppression of rotations about all axes except for the $\pm X$ axis in the large cat size limit $\alpha^2 \gg 1$. The limit $\alpha^2 \rightarrow 0$ recovers the familiar property $\matelphi \rightarrow \matelx \rightarrow 1$ of the bare qubit, where rotations about all axes are equally possible and there is no dependence on $\varphi$.

\subsection{Two-qubit gate}
The two-qubit gate (qubits $i = A, B$) we consider is based on the tunable beamsplitter interaction
\begin{equation}
    g(t) \qty(e^{i \tilde{\varphi}} a_A^\dagger a_B + e^{-i \tilde{\varphi}} a_B^\dagger a_A) = g(t) \qty(\cos(\tilde{\varphi}) \qty(a_A^\dagger a_B + a_B^\dagger a_A) + i \sin(\tilde{\varphi}) \qty(a_A^\dagger a_B - a_B^\dagger a_A)).
\end{equation}
Projecting the ladder operators to the computational subspace of both qubits ($a_i = (\matelx^i X_i + i \mately^i Y_i) / 2$), similar to the previous section, we obtain the effective drive
\begin{equation}
    \frac{g(t)}{2} \qty(\cos(\tilde{\varphi}) \qty(\matelx^A \matelx^B X_A X_B + \mately^A \mately^B Y_A Y_B) + i \sin(\tilde{\varphi}) \qty(\matelx^A \mately^B X_A Y_B - \mately^A \matelx^B Y_A X_B)).
\end{equation}
Depending on the drive phase $\tilde{\varphi}$, different rotation axes can be accessed. Note that only the $XX$ term does not contain a suppressed factor, as each other term contains at least one factor $\mately^i \propto e^{-2\alpha_i^2}$, significantly reducing the achievable Rabi rate of rotations about those axes. Hence we simplified the discussion to $\tilde{\varphi} = 0$ in the main text.
\section{Details on single-qubit gates}
In this work most pulse shapes are given by scaled versions of the truncated Gaussian function that was introduced in Ref.~\cite{xuEngineeringFastBiaspreserving2022}:
\begin{equation}
    \label{eq:trunc_gauss}
    f(t) = \qty[\frac{e^{-(t-T/2)^2/2\sigma^2} - e^{-(T/2)^2/2\sigma^2}}{1 - e^{-(T/2)^2/2\sigma^2}}]^m = \qty[\frac{e^{-(t/T-1/2)^2/2} - e^{-1/8}}{1 - e^{-1/8}}]^2.
\end{equation}
The last equation holds for our choice of parameters $\sigma = T$ and $m = 2$, which were also used in Ref.~\cite{xuEngineeringFastBiaspreserving2022}. We use this shape for entire  pulses (e.g. $\epsp{}(t) = \varepsilon_0 f(t)$ for $t \in [0, T]$) as well as ramps (e.g., ramp up: $\epsp{}(t) = \varepsilon_0 f(t)$ for $t \in [0, \tau]$, ramp down: $\epsp{}(t) = \varepsilon_0 f(t+\tau)$ for $t \in [0, \tau]$). 

The pulse schemes we present in the manuscript are characterized by specific optimizable parameters, which often correspond to pulse amplitudes $\varepsilon_0$ or ramp times $\tau$; we explicitly state the relevant parameters in the main text for each scheme. For example, the $Y$ gate pulse, shown in Fig.~2(d), has the free parameters $\epspzero{y}$ (the $Y$-component amplitude of the single-photon drive) and $\epspptilzero$ (the amplitude of the two-photon drive modulation). The DRAG drive $\epsp{x}(t)$ is fully determined by the pulse shapes $\epsp{y}(t)$ and $\epspptil(t)$, which is derived in Section~\ref{sec:ygate} of this SI. Generally, for a pulse scheme described by a set of parameters $\gamma_i$, we find the optimal choice of parameters for each data point in our robustness studies by minimizing the average infidelity over the detuning range:
\begin{equation}
    \label{eq:pulse_opt}
    \hat{\gamma}_i = \arg \min_{\gamma_i} \infidavg(T; \gamma_i) = \arg \min_{\gamma_i} \frac{1}{2\Delta_\mathrm{max}} \int_{-\Delta_\mathrm{max}}^{\Delta_\mathrm{max}} \dd{\Delta} \infid(T, \Delta; \gamma_i).
\end{equation}
Here $\infid(T, \Delta; \gamma_i)$ denotes the infidelity introduced in Eq.~(4) of the main text, where the pulse that implements the gate $U(T, \Delta)$ is parameterized by $\gamma_i$. The integral in Eq.~\eqref{eq:pulse_opt} is evaluated numerically by computing the infidelity at eleven equally spaced points in the interval $\qty[-\Delta_\mathrm{max}, \Delta_\mathrm{max}]$ and applying the Simpson rule for a better approximation. We perform the optimization exhaustively, which is feasible for the low-dimensional pulse parameterization we apply. 

\subsection{$X(\theta)$ rotations}

For this well-known case we restrict ourselves to a simple gate protocol with pulse shape $\epsilon_x(t) = \epsilon_{x, 0} f(t)$, where the amplitude $\epsilon_{x, 0}$ is the only free pulse parameter and optimized according to the method outlined in the previous section. We see in Fig.~2(c) that the average infidelity increases quickly in the regime of small gate time $T$ and cat size $\alpha^2$. This is because the required drive amplitudes become large while the gap to leakage states is rather small ($\simeq 2 \alpha^2 K$), leading to off-resonant excitation. A more sophisticated pulse design or the introduction of a DRAG component (see discussion in Section~\ref{sec:ygate}) could mitigate this effect, however in our work we do not optimize the $X$ protocol further as the main focus lies with the $Z$ rotations. For larger cat sizes the leakage issue does not emerge because the gap to to leakage states grows linearly with $\alpha^2$ and the drive amplitude required for the gate decreases since $\matelx \simeq 2 \alpha$ increases.
\subsection{$Y(\theta)$ rotations}
\label{sec:ygate}
Rotations about the $Y$ axis are exponentially suppressed in the cat size as $\mately \simeq 2 \alpha e^{-2 \alpha^2}$, which is why we propose the temporary decrease of the two-photon pump, $\epspp(t) = \epsppzero + \epspptil(t)$ with $\epspptil(t)/K \leq 0$, to lower the cat size $\alpha^2(t) = \epspp(t)/K$ and increase $\mately(t)$. In our protocol the cat size ramp $\epspptil(t)$ happens simultaneously with the actual gate drive $\varepsilon(t)$ to form a rotation gate $Y(\theta)$ with duration $T$. 

A small matrix element $\mately$ necessitates a large drive component $\epsp{y}(t)$, which leads to the undesired off-resonant coupling to states outside the computational subspace. In particular, we observe Stark shifts of the computational states $\ket{0}$ and $\ket{1}$, causing a tilted rotation axis $\hat{n}$ with a nonzero $n_z$ component. We aim to cancel this effect by introducing a DRAG (Derivate Removal by Adiabatic Gate \cite{motzoiSimplePulsesElimination2009,gambettaAnalyticControlMethods2011}) term $\epsp{x}(t)$ in the $X$ quadrature, which we derive in the following. The system Hamiltonian reads
\begin{align}
    \ham(t) &= -\frac{K}{2} a^{\dagger 2} a^2 + \frac{\epsppzero + \epspptil(t)}{2} \qty(a^2 + a^{\dagger 2}) + \frac{\epsp{y}(t)}{2} (-i) \qty(a - a^\dagger) + \frac{\epsp{x}(t)}{2} \qty(a + a^\dagger) \\
    &= \ham_0 + \frac{\epspptil(t)}{2} \ham_2 + \frac{\epsp{y}(t)}{2} \ham_y + \frac{\epsp{x}(t)}{2} \ham_x \\
    &= \ham_{02y}(t) + \frac{\epsp{x}(t)}{2} \ham_x.
\end{align}
Defining the instantaneous eigenstates $\ket{\psi_{j}(t)}$ with respect to the main drive Hamiltonian $\ham_{02y}$ via $\ham_{02y}(t) \ket{\psi_{j}(t)} = \tilde{E}_j(t) \ket{\psi_{j}(t)}$, we can conclude for the time evolution of an arbitrary state $\ket{\psi(t)} = \sum_j c_j(t) \ket{\psi_j(t)}$:
\begin{align}
    \label{eq:y_adse}
    \dot{c}_j(t) = -i \tilde{E}_j(t) c_j(t) - \sum_k c_k \qty(i \frac{\epsp{x}(t)}{2} \mel{\psi_j(t)}{\ham_x}{\psi_k(t)} + \ip*{\psi_j(t)}{\dot{\psi}_k(t)}).
\end{align}
All controls start at 0, therefore we have $\ket{\psi_0(0)} = \ket{+i} = (\ket{0} + i \ket{1})/\sqrt{2}$ and $\ket{\psi_1(0)} = \ket{-i} = (\ket{0} - i \ket{1})/\sqrt{2}$ due to the presence of $\ham_y$, where $\ket{0/1} = \ket{C_{\alpha}^\pm}$ and $\alpha^2 = \epsppzero/K$. In the ideal case the desired $Y$ rotation in the computational basis is realized through the accrual of a phase difference between $\ket{\psi_0(t)}$ and $\ket{\psi_1(t)}$ without any transitions in the process. However, the sum in Eq.~\eqref{eq:y_adse} includes terms that describe such transitions between instantaneous eigenstates, and using the DRAG control $\epsp{x}(t)$ we aim to specifically cancel the $\ket{\psi_0(t)} \leftrightarrow \ket{\psi_1(t)}$ transition. Thus we require
\begin{equation}
    \epsp{x}(t) = \frac{2i \ip*{\psi_1(t)}{\dot{\psi}_0(t)}}{\mel{\psi_1(t)}{\ham_x}{\psi_0(t)}}.
\end{equation}
In order to rewrite $\ip*{\psi_1(t)}{\dot{\psi}_0(t)}$, we consider
\begin{align}
    \partial_t \qty(\ham_{02y}(t) \ket{\psi_0(t)}) &= \partial_t \qty(\tilde{E}_0(t) \ket{\psi_0(t)}) \\
    \dot{\ham}_{02y}(t) \ket{\psi_0(t)} + \ham_{02y}(t) \ket*{\dot{\psi}_0(t)} &= \dot{\tilde{E}}_0(t) \ket{\psi_0(t)} + \tilde{E}_0(t) \ket*{\dot{\psi}_0(t)} \\
    \implies \ip*{\psi_1(t)}{\dot{\psi}_0(t)} &= \frac{\mel{\psi_1(t)}{\dot{\ham}_{02y}(t)}{\psi_0(t)}}{\tilde{E}_0(t) - \tilde{E}_1(t)} \\
    &= \frac{\dot{\tilde{\varepsilon}}_2(t) \mel{\psi_1(t)}{\ham_2}{\psi_0(t)} + \dot{\varepsilon}_y(t) \mel{\psi_1(t)}{\ham_y}{\psi_0(t)}}{2(\tilde{E}_0(t) - \tilde{E}_1(t))}.
\end{align}
Therefore the full DRAG correction reads
\begin{equation}
    \label{eq:drag}
    \epsp{x}(t) = \frac{i}{\tilde{E}_0(t) - \tilde{E}_1(t)} \frac{\dot{\tilde{\varepsilon}}_2(t) \mel{\psi_1(t)}{\ham_2}{\psi_0(t)} + \dot{\varepsilon}_y(t) \mel{\psi_1(t)}{\ham_y}{\psi_0(t)}}{\mel{\psi_1(t)}{\ham_x}{\psi_0(t)}}.
\end{equation}

This is the exact solution that cancels the desired transition at every point in time $t$. It can be simplified by treating the control $\epsp{y}(t)$ perturbatively and approximating the energies and matrix elements. Expanding the numerator and denominator to first order ultimately yields
\begin{equation}
    \label{eq:drag_approx}
    \epsp{x}(t) = \frac{1}{4 \mately(t) \matelx(t)} \qty(\frac{\mately^{1 \rightarrow 2}(t)^2}{E_2(t)} - \frac{\mately^{0 \rightarrow 3}(t)^2}{E_3(t)}) \dot{\varepsilon}_y(t),
\end{equation}
where $\mately^{0 \rightarrow 3}(t) = \abs{\!\mel{3}{H_y}{0}(t)}$ and $\mately^{1 \rightarrow 2}(t) = \abs{\!\mel{2}{H_y}{1}(t)}$ describe the coupling of the computational states to the first two excited states $\ket{2}$ (energy $E_2(t)$) and $\ket{3}$ (energy $E_3(t)$) with respect to the instantaneous KC of size $\alpha^2(t)$. Eq.~\eqref{eq:drag_approx} recovers the familiar structure where the DRAG drive $\epsp{x}(t)$ is proportional to the derivative of the primary drive $\epsp{y}(t)$, here for the case of two relevant transitions, which was also discussed in Ref.~\cite{gambettaAnalyticControlMethods2011}.

\begin{figure}[htbp]
    \centering
    \includegraphics[scale=1.]{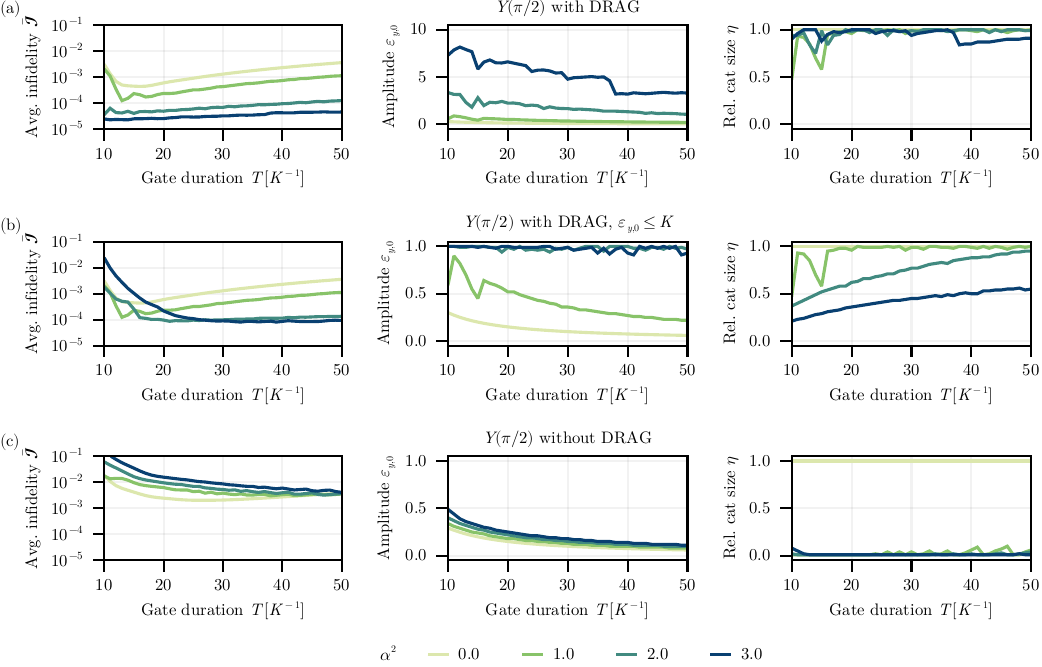}
    \caption{Optimization results of the $Y(\pi/2)$ gate for different settings, showing average infidelity $\infidavg$, drive amplitude $\epspzero{y}$ and relative cat size $\eta = \alpha^2(T/2)/\alpha^2$ at the most ramped-down point. (a) Gate scheme including the DRAG correction $\epsp{x}$. This is the main result, which is also included in Fig.~2 of the manuscript. The cat size is barely ramped down to retain the KC protection, however this makes large drive amplitudes necessary, which can become a limiting factor. (b)  When constraining the drive amplitude, the cat size has to be ramped increasingly more for larger initial $\alpha^2$ in order to counteract the suppressed $\mately$ element. Sacrificing the KC protection limits the achievable average infidelity. (c) Without the implementation of a DRAG correction, the coupling to excited states combined with the suppressed $\mately$ element leads to a Stark shift and ultimately to a loss in fidelity, hence the found solution is to ramp the cat size down all the way to zero. No robustness improvement over the bare qubit ($\alpha^2 = 0$) is achieved.}
    \label{fig:drag}
\end{figure}

We perform the robustness study for the $Y(\pi/2)$ gate and compare different settings. In all cases, the drive's $Y$-component $\epsp{y}(t) = \epspzero{y} f(t)$ as well as the two-photon ramp $\epspptil(t) = \epspptilzero f(t)$ are given by truncated Gaussians, where the amplitudes $\epspzero{y}$ and $\epspptilzero$ are optimized over. We include the results in Fig.~\ref{fig:drag}, which shows the optimized average infidelities $\infidavg$ alongside the optimized pulse parameters for the different settings, where instead of $\epspptilzero$ we plot the relative cat size $\eta = \alpha^2(T/2)/\alpha^2 = 1 + \epspptilzero/(\alpha^2 K)$ at the most ramped-down point ($t = T/2$).

Fig.~\ref{fig:drag}(a) reiterates the results from the main text for the $Y(\pi/2)$ gate with DRAG correction and further shows the pulse parameters. We note that under the assumption of an unconstrained drive power supply, the optimal solution is to keep the cat size high in order to retain as much KC protection as possible and achieve improving infidelities as the cat size grows. However, this necessitates the application of a rapidly growing drive strenght to compensate for the exponentially suppressed matrix element $\mately$. Therefore, as long as the required power can be provided, this approach is feasible and can produce average infidelities well below $\infidavg < 10^{-4}$ for cat sizes $\alpha^2 \gtrsim 2$ with short gate times.

As the supply of drive power can be a physical limitation, we consider the case where the dominant $Y$-component is upper bounded, and for this study we choose $\epspzero{y} \leq K$ as an example. Fig.~\ref{eq:drag}(b) shows how the robustness and pulse parameters are affected by this constraint. While there are no changes for the small cat sizes $\alpha^2 \lesssim 1$ as these system do not require that much drive power in the unconstrained case, we observe a clear loss of robustness for the larger cat sizes we consider, especially in the short gate time regime. Due to the upper limit on the drive amplitude, the cat size has to be ramped down significantly during the gate protocol to increase the matrix element $\mately$ and speed up the gate angle accumulation. The associated reduction of KC protection leads to an increase of the average infidelity, and we essentially find no improvement when increasing the cat size beyond $\alpha^2 \gtrsim 2$.

In order to emphasize the importance of the DRAG correction, we include in Fig.~\ref{fig:drag}(c) the (unconstrained) gate optimization without a DRAG term $\epsp{x}(t)$. We observe that in this setting the best performing solution corresponds to ramping the KC down all the way to the bare qubit at $\alpha^2 = 0$ (equivalent to $\eta = 0$) to realize the rotation. This is because at nonzero cat size the KC suffers from a Stark shift that emerges from the strong drive (due to exponentially suppressed matrix element $\mately$) coupling to the excited states, which is not suppressed with the cat size. This coherent error hurts the gate fidelity more than the influence of the frequency noise $\Delta$, hence the reduction to the bare qubit and no robustness improvement is achieved by making the KC larger. Therefore, the addition of the DRAG correction is absolutely crucial. 
\subsection{$Z(\theta)$ rotations}
\label{sec:zgate}
In many quantum computing platforms, rotations about the $Z$ axis are implemented ``virtually'' by adjusting the phase $\tilde{\varphi}$ of the single-photon drive for subsequent gates \cite{mckayEfficientGatesQuantum2017}. However, this relies on the assumption that a change in $\tilde{\varphi}$ induces a significant change in the azimuthal angle $\varphi$ of the rotation axis, which does not apply in the case of the KC due to the exponential suppression of matrix elements $\matelphi$ for $\sin(\varphi) \neq 0$ (see Section~\ref{sec:1p}). This is why $Z(\theta)$ gates have to be implemented physically (i.e., with an actual gate pulse) beyond a certain cat size as the suppression becomes too dominant, and therefore we consider $\alpha^2 \gtrsim 1$ for this gate scheme.

In this section we expand on our $Z(\theta)$ protocol \Circled{1}$\leftrightarrow$\Circled{2} and provide reasoning for the emergence of robustness using a simplified model of the adiabatic evolution. This further motivates an alternative gate scheme \Circled{1}$\leftrightarrow$\Circled{2'} that does not follow the robust line (see Fig.~3(a) in the main text). We start with the Hamiltonian
\begin{equation}
    \ham(t) = \delta(t) a^\dagger a - \frac{K}{2} a^{\dagger 2} a^2 + \frac{\epspp(t)}{2} \qty(a^2 + a^{\dagger 2}),
\end{equation}
where we assume control over the detuning $\delta(t)$ and, like for $Y(\theta)$ gates discussed in Section~\ref{sec:ygate}, the cat size $\alpha^2(t) = \epspp(t)/K$. Considering the evolution of a quantum state $\ket{\psi(t)} = \sum_j c_j(t) \ket{\psi_j(t)}$ expressed in the instantaneous eigenbasis defined by $\ham(t) \ket{\psi_j(t)} = E_j(t) \ket{\psi_j(t)}$, we obtain the equation of motion
\begin{equation}
    \dot{c}_j(t) = -i E_j(t) c_j(t) - \sum_k c_k(t) \ip*{\psi_j(t)}{\dot{\psi}_k(t)}.
\end{equation}
We define the gauge such that $\arg(c_j(0)) = 0$. In our idealistic and simplified analysis of an adiabatic evolution, where we assume no transitions and geometric phases, we can disregard the right term and only consider the phase accumulation of each eigenstate $\ket{\psi_j(t)}$ due to its energy $E_j(t)$:
\begin{equation}
    c_j(t) = \exp(-i \int_0^t \dd{t'} E_j(t')) c_j(0).
\end{equation}
The gate schemes we consider start and end with the non-detuned KC ($\delta = 0$, cat size $\alpha^2$, point \Circled{1}) and then trace the parameter space ($\delta(t)$, $\alpha^2(t)$) approximately adiabatically in a time $T$. The number operator $a^\dagger a$ breaks the KC's degeneracy in the computational basis $\qty{\catp, \catm}$ basis, therefore we have $\ket{\psi_0(0)} = \ket{\psi_0(T)} = \catp$ and $\ket{\psi_1(0)} = \ket{\psi_1(T)} = \catm$. This is why the relative phase accrued between the instantaneous eigenstates $\ket{\psi_0(t)}$ and $\ket{\psi_1(t)}$ leads to a $Z(\theta)$ rotation in the computational basis, with rotation angle
\begin{equation}
    \theta = \arg\qty(\frac{c_1(T)}{c_0(T)}) = - \int_0^T \dd{t} E_{01}(t). 
\end{equation} 
Here $E_{01}(t) = E_0(t) - E_1(t) = E_{01}(\delta(t), \alpha^2(t))$ denotes the instantaneous energy gap along the trajectory ($\delta(t)$, $\alpha^2(t)$). In the detuning range $\delta \in [0, K]$ we consider in this work, the emerging energy gap is nonnegative, i.e. $E_{01}(t) \geq 0$, thus the rotation occurs clockwise and the gate angle $\theta$ grows in the negative direction.

Detuning errors $\Delta$ lead to shifts in the energy gap $E_{01}(t; \Delta)$ along the trajectory and therefore cause erronous rotation angle $\theta(\Delta)$. To first order in $\Delta$, we can write
\begin{align}
    \label{eq:theta_error}
    \theta(\Delta) = -\int_0^T \dd{t} E_{01}(\delta(t) + \Delta, \alpha(t)) &\simeq -\int_0^T \dd{t} \qty(E_{01}(\delta(t), \alpha(t)) + \Delta \, \pdv{E_{01}}{\delta} \,\!(\delta(t), \alpha(t))) \\
    \label{eq:theta_error_static}
    &= \theta(\Delta = 0) + \Delta \, \partial_\delta \theta(\Delta = 0) \\[5pt]
    \label{eq:dtheta_error}
    \implies \partial_\delta \theta(\Delta = 0) &= \int_0^T \dd{t} \pdv{E_{01}}{\delta} \,\!(\delta(t), \alpha(t)).
\end{align}
The robustness of our primary scheme \Circled{1}$\leftrightarrow$\Circled{2} originates from the fact that by design the energy gap derivative $\partial_\delta E_{01}$ is either vanishly small or exactly zero along the entire trajectory. During the inital detuning ramp $\delta = 0 \rightarrow \delta_\mathrm{rob}(\alpha^2)$ and final detuning ramp $\delta = \delta_\mathrm{rob}(\alpha^2) \rightarrow 0$ the gap derivative is exponentially small in $\alpha^2$ due to the KC protection, and while traversing the robust line $(\delta_\mathrm{rob}(\alpha^2(t)), \alpha^2(t))$ we have $\partial_\delta E_{01}(t) = 0$ by construction. This enables average gate infidelities $\infidavg < 10^{-5}$ as can be seen in Fig.~3(d) in the main text.

Since this gate scheme relies on adiabatic evolution, nonadiabatic errors are a relevant source of infidelity, in particular in the low duration regime. This reason for the spikes in that regime in Fig.~3(d) (main text).  More carefully designed pulse shapes could help mitigate this effect. Especially the two-photon pulse $\epspptil(t)$ should be looked at in this context, as the relevant spectral gaps, which critically influence the validity of the adiabatic approximation, strongly depend on the cat size. 

We note that Eqs.~\eqref{eq:theta_error_static} and \eqref{eq:dtheta_error} are only valid for static detuning errors as we take $\Delta$ out of the integral. However, the limitation to static errors is not strictly necessary for the scheme \Circled{1}$\leftrightarrow$\Circled{2}, because the error term under the integral in Eq.~\eqref{eq:theta_error} is small or zero for all $t$. This is why high robustness can be expected for certain dynamical errors $\Delta(t)$ as well, and we expand on the extension to time-dependent errors in Section~\ref{sec:time_errors}.

\subsubsection*{Alternative gate scheme \Circled{$1$}$\leftrightarrow$\Circled{$2'$}}
\label{sec:zgate_alternative}
If we restrict ourselves to static detuning errors $\Delta$, then a different gate scheme with a simpler implementation becomes possible. It allows to relax the constraint of $\partial_\delta E_{01}(t)$ (approximately) vanishing at all times $t$, and instead we can only require  $\partial_\delta \theta  (\Delta = 0) = 0$ over the entire gate evolution in order to have first-order robustness against static detuning errors. We see from Eq.~\eqref{eq:dtheta_error} that this corresponds to a vanishing \emph{time-averaged} energy gap. 

\begin{figure}[htbp]
    \centering
    \includegraphics[scale=1.0]{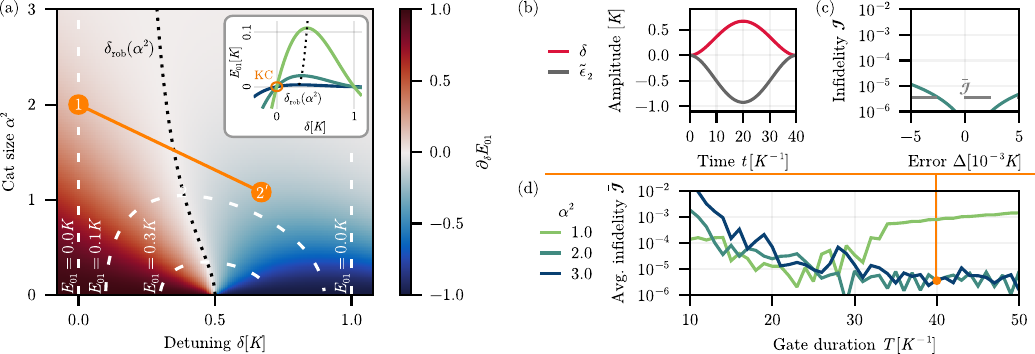}
    \caption{Alternative realization of the $Z(-\pi/2)$ gate using the \Circled{1}$\leftrightarrow$\Circled{2'} scheme. (a) The energy gap derivative $\partial_\delta{E_{01}}(\delta, \alpha^2)$ as a function of the detuning $\delta$ and cat size $\alpha^2$. The robust line $\deltarob(\alpha^2)$ (highlighted by a black dashed line) separates the parameter space into two regions with different signs of the energy gap derivative $\partial_\delta E_{01}(\delta, \alpha^2)$, which enables first-order insensitivity to static errors $\Delta$ on averaging over the trajectory. An example trajectory is drawn with an orange solid line. The inset shows the energy spectrum. (b) Control pulse for the solid trajectory shown in (a), which simultaneously tunes the detuning $\delta$ and the two-photon pump strength via $\epspptil$. (c) The infidelity of the pulse (b) when simulated over a range of detuning errors $\Delta$. (d) The average infidelity $\infidavg$ of pulses of type (b) for different durations and cat sizes.}
    \label{fig:zgate_alternative}
\end{figure}

The insensitive line $\deltarob(\alpha^2)$ traces the maxima of the energy gap and therefore separates parameter space into two parts, which is visualized in Fig.~\ref{fig:zgate_alternative}(a). On the red (left) side we have $\partial_\delta E_{01}(\delta, \alpha^2) > 0$, while on the blue (right) side we have $\partial_\delta E_{01}(\delta, \alpha^2) < 0$. Therefore, there are many gate trajectories that lead to first-order robustness because they start with a non-detuned KC on the left side and cross over to the right side before returning to the inital point on the left side, which, when pulse parameters are chosen properly, results in a vanishing time-averaged energy gap. 

To illustrate this scheme, we consider the subset of such gate trajectories that traverse the parameter space in a straight line from a inital KC ($\delta = 0$, cat size $\alpha^2$) at point \Circled{1} to the midpoint \Circled{2'} and back to \Circled{1}. The midpoint \Circled{2'} with parameters $(\delta_\mathrm{max}, \alpha'^2)$ may lie anywhere in parameter space for this scheme, however, as discussed above, robustness is only possible when it is located on the right side of the insensitive line ($\delta_\mathrm{max} > \deltarob(\alpha'^2)$). An example trajectory is highlighted in Fig.~\ref{fig:zgate_alternative} with an orange line. We specifically choose the truncated Gaussian pulse shape $f(t)$ to trace the trajectory, so that we have:
\begin{align}
    \mqty(\delta(t) \\ \alpha^2(t)) &= f(t) \mqty(\delta_\mathrm{max} \\ \alpha'^2)  + (1 - f(t)) \mqty(0 \\ \alpha_1^2) \\[5pt]
    \implies \mqty(\delta(t) \\ \epspptil(t)) &= f(t) \mqty(\delta_\mathrm{max} \\ \epspptilzero)
\end{align} 
Here we introduced the two-photon pump difference $\epspptil(t)/K \leq 0$ again, which describes the change in cat size from the starting point via $\alpha^2(t) K = \epspp(t) = \epsppzero + \epspptil(t)$. The amplitudes $\delta_\mathrm{max}$ and $\epspptilzero$ are the optimizable parameters.

We perform the $Z(-\pi/2)$ robustness study for this \Circled{1}$\leftrightarrow$\Circled{2'}  scheme like we did it for the \Circled{1}$\leftrightarrow$\Circled{2} in the main text and present the results in Figs.~\ref{fig:zgate_alternative}(b)-(d). We find highly robust pulses that achieve infidelities well below $10^{-5}$ for all cat sizes studied. Similarly to the primary scheme, we note higher infidelities in the low duration regime with stronger spikes due to nonadiabatic errors. The infidelity for $\alpha^2 = 1$ grows significantly for $T \gtrsim 30 \, K^{-1}$ because feasible trajectories that implement the right rotation angle $\theta = \pi/2$ do no extend far enough into the blue (right) region (as most of the phase is accumulated in the red (left) region) to average out the energy gap $\partial_\delta E_{01}$.

We observe small infidelities over a greater range of gate times $T$ compared to the primary scheme, which is due to the greater freedom in pulse trajectories and simpler pulse shapes. Regarding the latter, we emphasize that this scheme \Circled{1}$\leftrightarrow$\Circled{2'} is more practical and easier to implement compared to \Circled{1}$\leftrightarrow$\Circled{2} because it only uses analytical pulse shapes (truncated Gaussians) that can be easily calibrated in experiment (amplitude tuning). The scheme \Circled{1}$\leftrightarrow$\Circled{2} is more challenging to realize in this regard as the computation of the robust line $\deltarob(\alpha^2)$ requires numerical diagonalization of the Hamiltonian in theory and a more involved calibration in practice. However, it is important to keep in mind that this simpler nature of the \Circled{1}$\leftrightarrow$\Circled{2'} scheme comes at the cost of being effective to static errors $\Delta$ only.
\section{Discussion about time-dependent errors $\Delta(t)$}
\label{sec:time_errors}
While we focused on static frequency shifts $\Delta$ in our work, here we give qualitative and quantitative reasoning why we can expect our KC framework to be robust to (well-behaved) time-dependent shifts $\Delta(t)$ as well. In the case of idling, $X$ and $XX$ rotations, where we have the full KC protection related to the cat size $\alpha^2$, this is easy to see because of the suppressed energy gap between the computational states:
\begin{equation}
    E_{01}(\Delta(t), \alpha^2) \simeq \partial_\delta E_{01}(0, \alpha^2) \, \Delta(t).
\end{equation}
This holds for slow-varying perturbations $\Delta(t)$ where we can consider the instantaneous energy spectrum of the system. More generally, we can look at the action of the perturbation operator (the photon number operator in this case) in the computational subspace
\begin{equation}
    a^\dagger a \simeq \alpha^2 \mathbb{1} - 2 \alpha^2 e^{-2 \alpha^2} Z
\end{equation}
and note the exponential suppression of phase errors ($\sim Z$).

In the following we focus the discussion on our adiabatic $Z(\theta)$ schemes, where we assume slowly-varying perturbations $\Delta(t)$ so we can apply the adiabatic modeling we already introduced in Section~\ref{sec:zgate}. Employing the same derivation that lead to Eq.~\eqref{eq:theta_error}, we can approximate the error of the gate angle due to the shift $\Delta(t)$ as
\begin{equation}
    \label{eq:theta_til_dyn}
    \tilde{\theta}[\Delta] = \theta[\Delta] - \theta_\mathrm{id} \simeq - \int_0^T \Delta(t) \partial_\delta E_{01}(t) \dd{t},
\end{equation}
where $\partial_\delta E_{01}(t) = \partial_\delta E_{01}(\delta(t), \alpha^2(t))$ denotes the energy gap derivative along the trajectory. We use the notation $g[\Delta]$ to indicate that $g$ is a functional that maps the function $\Delta(t)$ to a scalar. The gate infidelity is given by
\begin{equation}
    \infid[\Delta] = 1 - \frac{1}{4} \abs{\Tr[Z\qty(\tilde{\theta}[\Delta])]}^2 = \sin^2\qty(\frac{\tilde{\theta}[\Delta]}{2}) \simeq \frac{\tilde{\theta}[\Delta]^2}{4} = \frac{1}{4} \int_0^T \dd{t_1} \int_0^T \dd{t_2} \Delta(t_1) \Delta(t_2) \underbrace{\partial_\delta E_{01}(t_1) \partial_\delta E_{01}(t_2)}_{\equiv F(t_1, t_2)}.
\end{equation}

Ultimately we are interested in the average fidelity over different realizations of the frequency noise $\Delta(t)$,
\begin{equation}
    \label{eq:infidavg_dyn}
    \infidavg \simeq \frac{1}{4} \int_0^T \dd{t_1} \int_0^T \dd{t_2} \expval{\Delta(t_1) \Delta(t_2)} F(t_1, t_2),
\end{equation}
where we have introduced the correlation of $\Delta(t)$:
\begin{equation}
    \expval{\Delta(t_1) \Delta(t_2)} = \int \mathcal{D}[\Delta] \mathcal{P}[\Delta] \Delta(t_1) \Delta(t_2).
\end{equation}
with the probability distribution $\mathcal{P}[\Delta]$ and integration measure $\mathcal{D}[\Delta]$. We make the additional assumption that the frequency noise has no memory effect, i.e. the autocorrelation is a function of the time difference only, i.e. $\expval{\Delta(t_1) \Delta(t_2)} = \expval{\Delta(t_1 - t_2) \Delta(0)}$. This lets us rewrite Eq.~\eqref{eq:infidavg_dyn}:
\begin{align}
    \infidavg &\simeq \frac{1}{4} \int_0^T \dd{t_1} \int_0^T \dd{t_2} \expval{\Delta(t_1 - t_2) \Delta(0)} F(t_1, t_2) \\
    \label{eq:infidavg_dyn_dirac}
    &= \frac{1}{4} \int_0^T \dd{t_1} \int_0^T \dd{t_2} \int \dd{t} \hat{\delta}(t - (t_1 - t_2)) \expval{\Delta(t) \Delta(0)} F(t_1, t_2) \\
    \label{eq:infidavg_dyn_dirac_exp}
    &= \frac{1}{4} \int_0^T \dd{t_1} \int_0^T \dd{t_2} \int \dd{t} \int \frac{\dd{\omega}}{2\pi} e^{-i \omega (t - (t_1 - t_2))} \expval{\Delta(t) \Delta(0)} F(t_1, t_2) \\
    &= \int \frac{\dd{\omega}}{2\pi} \qty(\int \dd{t} e^{-i \omega t} \expval{\Delta(t) \Delta(0)}) \qty(\frac{1}{4} \int_0^T \dd{t_1} \int_0^T \dd{t_2} e^{i \omega (t_1 - t_2)} F(t_1, t_2)) \\
    \label{eq:infidavg_dyn_spectral}
    &= \int \frac{\dd{\omega}}{2\pi} S_\Delta(\omega) W(\omega).
\end{align}
In Eq.~\eqref{eq:infidavg_dyn_dirac} we inserted a Dirac-$\hat{\delta}$ distribution, which was then expanded to its exponential form in Eq.~\eqref{eq:infidavg_dyn_dirac_exp}. In its final form the average infidelity is expressed as a convolution of the power spectral density $S_\Delta(\omega)$ of the frequency noise and the weighting function $W(\omega)$, which is solely determined by the energy gap derivative along the gate trajectory:
\begin{align}
    S_\Delta(\omega) &= \int \dd{t} e^{-i \omega t} \expval{\Delta(t) \Delta(0)}, \\
    W(\omega) &= \frac{1}{4} \int_0^T \dd{t_1} \int_0^T \dd{t_2} e^{i \omega (t_1 - t_2)} F(t_1, t_2) = \frac{1}{4} \int_0^T \dd{t_1} \int_0^T \dd{t_2} e^{i \omega (t_1 - t_2)} \partial_\delta E_{01}(t_1) \partial_\delta E_{01}(t_2).
\end{align}

This result supports our robustness analysis for the $Z(\theta)$ gate, which was discussed in Section~\ref{sec:zgate}. Our primary scheme \Circled{1}$\leftrightarrow$\Circled{2} is designed such that we have $\partial_\delta E_{01}(t) \simeq 0$ along the trajectory, therefore we have $W(\omega) \simeq 0$ for all $\omega$ and, due to Eq.~\eqref{eq:infidavg_dyn_spectral}, we can expect very small infidelities independent of structure ($\sim S_\Delta(\omega)$) of the frequency noise.

For the other scheme \Circled{1}$\leftrightarrow$\Circled{2'}, however, the weighting function $W(\omega)$ does not vanish for most values of $\omega$, which makes this scheme less robust to time-varying errors $\Delta(t)$. In the special case of static frequency shifts $\Delta$ the power spectral density simplifies to $S_\Delta(\omega) = 2\pi \hat{\delta}(\omega) \Delta^2$ and the average infidelity becomes $\infidavg \simeq \Delta^2 W(0)$. The value $W(0)$ is proportional to the square of the trajectory-averaged energy gap derivative, which can be zero for certain trajectories as discussed in Section~\ref{sec:zgate_alternative}, leading to the first-order robustness to static frequency shifts that we observed in the study (see Fig.~\ref{fig:zgate_alternative}).

All in all we note that good robustness properties for a universal set of gates (in particular, $X(\theta)$, $XX(\theta)$, $Z(\theta)$) can be expected in the presence of time-dependent frequency shifts $\Delta(t)$ as well. However, it is important to keep in mind that the above analysis relied on an idealized adiabatic model and the assumption of slow-varying noise, which is helpful to gain intuition for the origin of robustness in the KC, but it might not hold for highly oscillating or less well-behaved noise.



\end{document}